\documentclass[showpacs,draft,preprint,floatfix]{revtex4}
\usepackage{amsfonts}

\usepackage{graphicx}
\usepackage{latexsym}
\usepackage{amsmath}

\begin{document}

\title{Scheme to measure squeezing and phase properties of a harmonic oscillator}

\author{G.T. Rub\'{\i}n-Linares and H. Moya-Cessa}
\affiliation{INAOE, Coordinaci\'on de Optica, Apdo. Postal 51 y
216, 72000 Puebla, Pue., Mexico}
\begin{abstract}
We propose a simple scheme to measure squeezing and phase
properties of a harmonic oscillator. We treat in particular the
case of a the field, but the scheme may be easily realized in ion
traps. It is based on integral transforms of measured atomic
properties as atoms exit a cavity. We show that by measuring
atomic polarizations it is possible, after a given integration, to
measure several properties of the field.
\end{abstract}
 \maketitle

The reconstruction of a quantum state is a central topic in
quantum optics and related fields \cite{1,2}. During the past
years, several techniques have been developed to achieve such
quantum measurements, many of them having as a basic tool quantum
tomography. Classical tomography is a method for building up a
picture of a hidden object using various observations from
different angles. In quantum optics, tomography has been applied
experimentally to reconstruct the quantum state of light from a
complete set of measured quantities, examples may be the direct
sampling of the density matrix of a signal mode in multiport
optical homodyne tomography \cite{3} and  tomographic
reconstruction by unbalanced homodyning \cite{4}. There have also
been proposals to measure electromagnetic fields inside cavities
\cite{5,6,7} which may be achieved through a finite set of
selective measurements of atomic states \cite{5,6} that make it
possible to reconstruct quasiprobability distribution functions
such as the Wigner function, that constitute an alternative
representation of a quantum state of the field.

Recently it was proposed a method to {\it measure} the Wigner
function of a quantized field in a cavity by  using a Fresnel
transformation of the atomic inversion of atoms traversing the
cavity \cite{5}. To measure the Wigner function is a difficult
task mainly because of the number of times the field has to be
prepared: to cover the complete phase space one has to displace
(in the ideal case, continuously) the initial state, i.e. the
initial field state has to be prepared a number of times. However
it may be that we need information only about some features of the
field, lt us say, squeezing properties, phase properties, etc.
Here we will show how squeezing properties and phase properties
may be measured by transforming data obtained after atoms pass
through the cavity and measuring them.

\subsection{Measuring squeezing}
To measure squeezing, we need to be able to measure quantities like
\begin{equation}
\langle \hat{X} \rangle = \langle \hat{a} \rangle+ c.c., \qquad \langle \hat{X}^2
\rangle = \langle \hat{a}^2 \rangle + \langle [\hat{a}^{\dagger}]^2\rangle+2\langle
\hat{n} \rangle + 1
\end{equation}
where $\hat{a}^{\dagger}$ and $\hat{a}$ are the creation and
annihilation operators for the field mode, respectively, obeying
$[\hat{a},\hat{a}^{\dagger}]=1$. Below we will show how it is
possible to measure quantities like $\langle \hat{a}^k \rangle,
k=1,2$ by utilizing an atom as a measuring device. We start by
writing the Hamiltonian of the two-level atom field resonant
interaction in the rotating wave approximation and the interaction
picture

\begin{equation}
\hat{H}=\lambda(\hat{a}^{\dagger}\hat{\sigma}_-+\hat{\sigma}_+\hat{a})= \lambda
\left(
\begin{array}{cc}
0 & \hat{a}
\\  \hat{a}^{\dagger}& 0
\end{array}
\right) \label{hamiltonian}
\end{equation}
where $\hat{\sigma}_+$ and $\hat{\sigma}_-$ are the raising and lowering atomic
operators, respectively. The atomic operators obey the commutation relation
$[\hat{\sigma}_+,\hat{\sigma}_-]=\hat{\sigma}_z$.

We can re-write Hamiltonian (\ref{hamiltonian}) with the help of
Susskind-Glogower operators \cite{susskind} as

\begin{equation}
\hat{H}=\lambda \hat{T} \left(
\begin{array}{cc}
0 & \sqrt{\hat{n}+1}
\\  \sqrt{\hat{n}+1}& 0
\end{array}
\right) \hat{T}^{\dagger}
\end{equation}
where
\begin{equation}
\hat{T} =  \left(
\begin{array}{cc}
1 & 0
\\ 0 & \hat{V}^{\dagger}
\end{array}
\right)
\end{equation}
where $\hat{n}=\hat{a}^{\dagger}\hat{a}$ and
$\hat{V}=\frac{1}{\sqrt{\hat{n}+1}}\hat{a}$. Note that
$\hat{T}^{\dagger} \hat{T}=1$ but $\hat{T}\hat{T}^{\dagger} \ne
1$. This allows us to write the evolution operator as
\begin{equation}
\hat{U}(t)= \hat{T} \left(
\begin{array}{cc}
\cos(\lambda t\sqrt{\hat{n}+1}) & -i\sin(\lambda t\sqrt{\hat{n}+1})
\\ -i\sin(\lambda t\sqrt{\hat{n}+1}) & \cos(\lambda t\sqrt{\hat{n}+1})
\end{array}
\right) \hat{T}^{\dagger} \label{evol}
\end{equation}
We are neglecting a term $|0\rangle\langle 0|$ in the above
evolution operator (in the element $22$), that however will not
affect the measurement of squeezing as we will consider the atom
in the excited state.  We consider the  field in an unknown state,
such that the initial state of the system is $|\psi(0)\rangle =
|e\rangle|\psi_F(0)\rangle$, the average of the operator
$\hat{\sigma}_+$ is given by
\begin{eqnarray}
\langle  \hat{\sigma}_+ \rangle &=& -i\langle \psi_F(0)|\cos(\lambda
t\sqrt{\hat{n}+1})\hat{V}^{\dagger}\sin(\lambda t\sqrt{\hat{n}+1})|\psi_F(0)\rangle
\\ \nonumber &=& -\frac{i}{2}\langle \psi_F(0)|\hat{V}^{\dagger}\left(\sin[\lambda
t\hat{\Delta}_+(\hat{n})] - \sin[\lambda
t\hat{\Delta}_-(\hat{n})]\right)|\psi_F(0)\rangle \label{sigma}
\end{eqnarray}
where \begin{equation} \hat{\Delta}_+(\hat{n})= \sqrt{\hat{n}+2}+\sqrt{\hat{n}+1},
\qquad \hat{\Delta}_-(\hat{n})= \sqrt{\hat{n}+2}-\sqrt{\hat{n}+1}.\end{equation} By
integrating (\ref{sigma}) by using a Fresnel integral \cite{5,gradshteyn}:

\begin{equation}
\int_0^{\infty}dT T \sin(T^2/A) \sin(BT) = \frac{AB}{4}\sqrt{\frac{\pi A}{2}}\left(
\cos\frac{AB^2}{4}+\sin\frac{AB^2}{4} \right)
\end{equation}
such that (with $\lambda t = T$)
\begin{eqnarray}
\int_0^{\infty}dT T \sin(T^2/A) \langle  \hat{\sigma}_+ \rangle =
-\frac{i}{2}\langle\psi_F(0)|
\hat{V}^{\dagger}(\hat{\gamma}_1-\hat{\gamma}_2)|\psi_F(0)\rangle \label{sigma2}
\end{eqnarray}
with
\begin{equation}
\hat{\gamma}_1^{(1)}  = \frac{A\hat{\Delta}_+(\hat{n})}{4} \sqrt{\frac{\pi
A}{2}}\left( \cos\left[ \frac{A\hat{\Delta}_+^2(\hat{n})}{4}\right]+\sin\left[
\frac{A\hat{\Delta}_+^2(\hat{n})}{4}\right]\right)
\end{equation}
and
\begin{equation}
\hat{\gamma}_2^{(1)} = \frac{A\hat{\Delta}_-(\hat{n})}{4}\sqrt{\frac{\pi
A}{2}}\left( \cos\left[ \frac{A\hat{\Delta}_-^2(\hat{n})}{4}\right]+\sin\left[
\frac{A\hat{\Delta}_-^2(\hat{n})}{4}\right]\right).
\end{equation}
Now we use the approximation $\sqrt{(\hat{n}+2)(\hat{n}+1)}\approx
\hat{n} +3/2$ that is valid for large photon numbers (see for
instance \cite{vogel}). We then can write
$\hat{\Delta}_+^2(\hat{n}) \approx 4\hat{n}+6$ and
$\hat{\Delta}_-^2(\hat{n}) \approx 0$. By setting $A=4\pi$ then we
obtain
\begin{equation}
\hat{\gamma}_1^{(1)} \approx (\sqrt{\hat{n}+2}+\sqrt{\hat{n}+1})\pi \cos\left[
(4\hat{n}+6)\pi\right] = \sqrt{2}\pi^2\hat{\Delta}_+(\hat{n})
\end{equation}
and
\begin{equation}
\hat{\gamma}_2^{(1)} \approx \sqrt{2}\pi^2
\hat{\Delta}_-(\hat{n}), \end{equation} so that the integral
transform (\ref{sigma2}) becomes
\begin{eqnarray}
\int_0^{\infty}dT T \sin(T^2/A) \langle  \hat{\sigma}_+ \rangle & = &
-i\sqrt{2}\pi^2 \langle\psi_F(0)| \hat{V}^{\dagger}\sqrt{\hat{n}+1}|\psi_F(0)\rangle
\\ \nonumber  & = & -i\sqrt{2}\pi^2 \langle\psi_F(0)|
\hat{a}^{\dagger}|\psi_F(0)\rangle . \label{sigmamas}
\end{eqnarray}

To measure $\langle\psi_F(0)| [\hat{a}^{\dagger}]^2|\psi_F(0)\rangle $ it is
necessary a two-photon transition. In this case,

\begin{equation}
\hat{H}=\lambda^{(2)} \hat{T}^2 \left(
\begin{array}{cc}
0 & \sqrt{(\hat{n}+1)(\hat{n}+2)}
\\  \sqrt{(\hat{n}+1)(\hat{n}+2)}& 0
\end{array}
\right) [\hat{T}^{\dagger}]^2
\end{equation}
where $\lambda^{(2)}$ is the interaction constant in the
two-photon case. One can find the evolution operator that will be
given by an expression similar to (\ref{evol}), just changing
$\sqrt{\hat{n}+1}\rightarrow\sqrt{(\hat{n}+1)(\hat{n}+2)}$,
$\hat{V}\rightarrow\hat{V}^2$ and
$\hat{V}^{\dagger}\rightarrow[\hat{V}^{\dagger}]^2$. It is then
easy to calculate the average of $\hat{\sigma}_+^{(2)}$,
 which is given by
\begin{eqnarray}
\langle  \hat{\sigma}_+^{(2)} \rangle &=& -i\langle
\psi_F(0)|\cos\left[\lambda^{(2)}
t\sqrt{(\hat{n}+1)(\hat{n}+2)}\right]
[\hat{V}^{\dagger}]^2\sin\left[\lambda^{(2)}
t\sqrt{(\hat{n}+1)(\hat{n}+2)}\right]|\psi_F(0)\rangle\nonumber
\\  &=& -\frac{i}{2}\langle \psi_F(0)|[\hat{V}^{\dagger}]^2\left(\sin[\lambda
t\hat{\delta}_+(\hat{n})] - \sin[\lambda
t\hat{\delta}_-(\hat{n})]\right)|\psi_F(0)\rangle
 \label{sigma2p}
\end{eqnarray}
with
\begin{equation}
\hat{\delta}_+(\hat{n})=
\sqrt{(\hat{n}+4)(\hat{n}+3)}+\sqrt{(\hat{n}+2)(\hat{n}+1)}\approx
2\hat{n}+5,
\end{equation} and
\begin{equation}
\hat{\delta}_-(\hat{n})=
\sqrt{(\hat{n}+4)(\hat{n}+3)}-\sqrt{(\hat{n}+2)(\hat{n}+1)}\approx
2.
\end{equation}
Again by (Fresnel) integration of the above expression
\begin{eqnarray}
\int_0^{\infty}dT T \sin(T^2/A) \langle  \hat{\sigma}_+ ^{(2)}\rangle  =  -i4\pi^2
\langle\psi_F(0)|
[\hat{V}^{\dagger}]^2(\hat{\gamma}_1^{(2)}-\hat{\gamma}_2^{(2)})|\psi_F(0)\rangle
 \label{sigmamas}
\end{eqnarray}
with
\begin{equation}
\hat{\gamma}_1^{(2)}  = \frac{A\hat{\delta}_+(\hat{n})}{4} \sqrt{\frac{\pi
A}{2}}\left( \cos\left[ \frac{A\hat{\delta}_+^2(\hat{n})}{4}\right]+\sin\left[
\frac{A\hat{\delta}_+^2(\hat{n})}{4}\right]\right)
\end{equation}
and
\begin{equation}
\hat{\gamma}_2^{(2)} = \frac{A\hat{\delta}_-(\hat{n})}{4}\sqrt{\frac{\pi
A}{2}}\left( \cos\left[ \frac{A\hat{\delta}_-^2(\hat{n})}{4}\right]+\sin\left[
\frac{A\hat{\delta}_-^2(\hat{n})}{4}\right]\right).
\end{equation}
By choosing the value  $A=8\pi$ we obtain
\begin{eqnarray}
\int_0^{\infty}dT T \sin(T^2/8\pi) \langle  \hat{\sigma}_+
^{(2)}\rangle & = &  -i8\pi^4 \langle\psi_F(0)|
[\hat{V}^{\dagger}]^2\sqrt{(\hat{n}+1)(\hat{n}+1)}|\psi_F(0)\rangle
\nonumber \\& =&  -i8\pi^4\langle [\hat{a}^{\dagger}]^2\rangle
 \label{sigmamas}
\end{eqnarray}

\subsection{Measuring phase properties}
Now we turn our attention to the phase properties of the field. The procedure,
although similar to the way of obtaining the quadratures of the field, will differ
in the integral forms that will be used. We compute now the average of
$\hat{\sigma}_+$ for the one-photon transition for an atom in the ground state and
the arbitrary field $|\psi_F(0)\rangle$, that reads
\begin{eqnarray}
\langle  \hat{\sigma}_+ \rangle = i\langle \psi_F(0)|\hat{V}^{\dagger}\sin(\lambda
t\sqrt{\hat{n}+1})\hat{V}^{\dagger}\cos(\lambda
t\sqrt{\hat{n}+1})\hat{V}|\psi_F(0)\rangle
\end{eqnarray}
that for a field with large number of photons may be approximated by
\begin{eqnarray}
\langle  \hat{\sigma}_+ \rangle &=& i\langle \psi_F(0)|\hat{V}^{\dagger}\sin(\lambda
t\sqrt{\hat{n}+1})\cos(\lambda
t\sqrt{\hat{n}})(1-|0\rangle\langle0|)|\psi_F(0)\rangle
\\ \nonumber &=& i\langle \psi_F(0)|\hat{V}^{\dagger}\sin(\lambda
t\sqrt{\hat{n}+1})\cos(\lambda t\sqrt{\hat{n}})|\psi_F(0)\rangle \label{phase}
\end{eqnarray}
Using the integral \cite{gradshteyn}
\begin{equation}
\int_0^{\infty} \frac{\sin(AT)\cos(BT)}{T} dT = \pi/2, \qquad
A>B>0 \label{inte}
\end{equation}
we can integrate (\ref{phase}) as:
\begin{eqnarray}
\int_0^{\infty} \frac{\langle  \hat{\sigma}_+ \rangle}{\lambda t} d(\lambda t) =
\frac{i\pi}{2}\langle \psi_F(0)|\hat{V}^{\dagger}|\psi_F(0)\rangle . \label{phase1}
\end{eqnarray}
To measure $\langle [\hat{V}^{\dagger}]^2\rangle$, again a
two-photon process is needed. The average value of $\langle
\hat{\sigma}_+\rangle$ with the atom entering in the ground state
is
\begin{eqnarray}
\langle  \hat{\sigma}_+^{(2)} \rangle &=& i\langle
\psi_F(0)|[\hat{V}^{\dagger}]^2\sin\left(\lambda^{(2)}
t\sqrt{(\hat{n}+1)(\hat{n}+2)}\right)[\hat{V}^{\dagger}]^2\cos\left(\lambda^{(2)}
t\sqrt{(\hat{n}+1)(\hat{n}+2)}\right)\hat{V}^2|\psi_F(0)\rangle
\nonumber
\\  &=& i\langle
\psi_F(0)|[\hat{V}^{\dagger}]^4\sin\left(\lambda^{(2)}
t\sqrt{(\hat{n}+3)(\hat{n}+4)}\right)\cos\left(\lambda^{(2)}
t\sqrt{(\hat{n}+1)(\hat{n}+2)}\right)\hat{V}^2|\psi_F(0)\rangle
\end{eqnarray}
by performing the integral
\begin{equation}
\int_0^{\infty} \frac{\langle  \hat{\sigma}_+^{(2)} \rangle}{T} dT
=\frac{i\pi}{2}\langle
\psi_F(0)|[\hat{V}^{\dagger}]^4\hat{V}^2|\psi_F(0)\rangle
\end{equation}
that for large intensity field approximates
\begin{equation}
\int_0^{\infty} \frac{\langle  \hat{\sigma}_+^{(2)} \rangle}{T} dT
=\frac{i\pi}{2}\langle
\psi_F(0)|[\hat{V}^{\dagger}]^2|\psi_F(0)\rangle
\end{equation}

Finally, the average value of $\hat{\sigma}_+$ may be found by
finding the average value of the observables $\hat{\sigma}_x$ and
$\hat{\sigma}_y$, with $\hat{\sigma}_+= \hat{\sigma}_x +
i\hat{\sigma}_y$. In order to find it, we write $\langle
\hat{\sigma}_x\rangle = Tr[\hat{\sigma}_x\hat{\rho}]=
Tr[\hat{R}\hat{\sigma}_z \hat{R}^{\dagger}\hat{\rho}] =
Tr[\hat{\sigma}_z \hat{R}^{\dagger}\hat{\rho}\hat{R}]=
Tr[\hat{\sigma}_z\hat{\rho}_R]$, i.e. the expectation value of
$\hat{\sigma}_z$ (the atomic inversion or the probability of
finding the atom in the excited state minus the probability of
finding it in the ground state) for a {\it rotated} (in the atomic
basis) density matrix with $\hat{R} =
\exp[(\hat{\sigma}_--\hat{\sigma}_+)\pi/4]$. A similar procedure
may be done for the expression $\langle \hat{\sigma}_y\rangle $.
Experimentally we would need to send an atom through a cavity that
contains the field under study, then (properly) rotate it after it
exits the cavity and measure its energy. We would need an
experimental setup as shown in Fig. \ref{fig1squeez}.

\begin{figure} [hbt]
\caption{\label{fig1squeez} Experimental setup to measure $\langle
\hat{\sigma}_x\rangle $ or $\langle \hat{\sigma}_y\rangle $. }
\end{figure}


\begin{thebibliography} {99}
\bibitem{1} K. Vogel and H. Risken, Phys. Rev. A 40, 2847 (1989).

\bibitem{2} U.Leonhardt, Measuring the Quantum State of Light, (Cambridge,
Cambridge    University Press) 1997.

\bibitem{3}A. Zucchetti, W. Vogel, M. Tasche, and D.-G. Welsch, Phys.
Rev. A 54, 1678 (1996).

\bibitem{4} S. Wallentowitz and W. Vogel, Phys. Rev. A 53, 4528 (1996).

\bibitem{5} P. Lougovsky, E. Solano, Z.M. Zhang, H. Walther, H. Mack, and
P.W. Schleich, Phys. Rev. Lett. 91, 010401(2003).

\bibitem{6} L.G. Lutterbach and L. Davidovich, Phys. Rev. Lett. 78, 2547
(1997).

\bibitem{7} H. Moya-Cessa, S.M. Dutra, J.A. Roversi, and A.
Vidiella-Barranco, J. of Mod. Optics  46, 555 (1999); H.
Moya-Cessa, J.A. Roversi, S.M. Dutra,  and A. Vidiella-Barranco,
Phys. Rev. A 60, 4029 (1999).

\bibitem{susskind} L. Susskind and J. Glogower, Physics {\bf 1}, 49 (1964).

\bibitem{gradshteyn} I.S. Gradshteyn and I.M. Ryzhik, {\it Table of integrals, series, and
products}, Academic Press, Inc., London 1980).

\bibitem{vogel} H. Moya-Cessa, S. Ch\'avez-Cerda and W. Vogel, J. Mod. Optics {\bf
66}(1998).
\end{thebibliography}
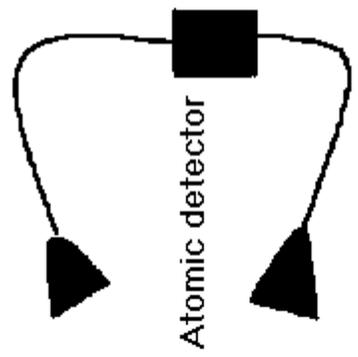
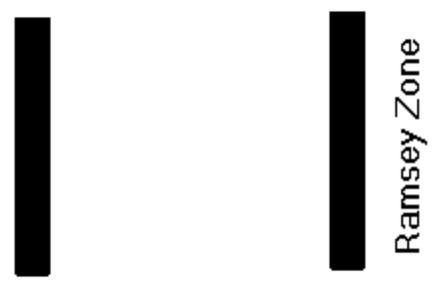
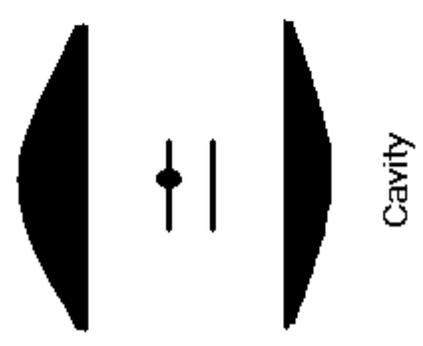
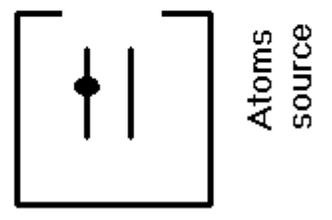
\end{document}